\documentclass[journal=nalefd,manuscript=letter]{achemso}
\usepackage{graphicx} % Required for inserting images
\usepackage{amsmath}
\usepackage[T1]{fontenc}
\usepackage{gensymb}
\usepackage{amssymb}

\author{Colin J. Riggert}
\affiliation[UMN]{School of Physics and Astronomy, University of
Minnesota, Minneapolis, Minnesota 55455, USA}
\author{Pim Lueb}
\affiliation[TUe]{Eindhoven University of Technology, Eindhoven, North Brabant 5600, The Netherlands}
\author{Tyler Littmann}
\affiliation[UMN]{School of Physics and Astronomy, University of
Minnesota, Minneapolis, Minnesota 55455, USA}

\author{Ghada Badawy}
\affiliation[TUe]{Eindhoven University of Technology, Eindhoven, North Brabant 5600, The Netherlands}
\author{Marco Rossi}
\affiliation[TUe]{Eindhoven University of Technology, Eindhoven, North Brabant 5600, The Netherlands}
\author{Paul A. Crowell}
\affiliation[UMN]{School of Physics and Astronomy, University of
Minnesota, Minneapolis, Minnesota 55455, USA}
\author{Erik P.A.M. Bakkers}
\affiliation[TUe]{Eindhoven University of Technology, Eindhoven, North Brabant 5600, The Netherlands}
\author{Vlad S. Pribiag}
\affiliation[UMN]{School of Physics and Astronomy, University of
Minnesota, Minneapolis, Minnesota 55455, USA}
\email{vpribiag@umn.edu}

\title{Few-Mode and Anisotropic Quantum Transport in InSb Nanoribbons Using an All-van der Waals Material-Based Gate}

\date{\today}

\begin{document}

\maketitle

\begin{abstract}
    High-quality electrostatic gating is a fundamental ingredient for successful semiconducting device physics, and a key element of realizing clean quantum transport. Inspired by the widespread improvement of transport quality when two-dimensional van der Waals (vdW) materials are gated exclusively by other vdW materials, we have developed a method for gating non-vdW materials with an all-vdW gate stack, consisting of a hexagonal boron nitride dielectric layer and a few-layer graphite gate electrode. We demonstrate this gating approach on MOVPE-grown InSb nanoribbons (NRs), a novel variant of the InSb nanowire, with a flattened cross-section. In our all-vdW gated NR devices we observe conductance features that are reproducible and have low- to near-zero gate hysteresis. We also report quantized conductance, which persists to lower magnetic fields and longer channel lengths than typical InSb nanowire devices reported to date. Additionally, we observe level splitting that is highly anisotropic in an applied magnetic field, which we attribute to the ribbon cross-section. The performance of our devices is consistent with the reduced disorder expected from the all-vdW gating scheme, and marks the first report of ballistic, few-modes quantum transport in a non-vdW material with an all-vdW gate. Our results establish all-vdW gating as a promising approach for high-quality gating of non-vdW materials for quantum transport, which is in principle applicable generically, beyond InSb systems. In addition, the work showcases the specific potential of all-vdW gate/InSb NR devices for enabling clean quantum devices that may be relevant for spintronics and topological superconductivity studies.
\end{abstract}

Control of the local electrostatic environment with gating is a cornerstone of semiconducting device physics, essential to reaching the ballistic, phase-coherent, few-mode, and few-electron regimes which define quantum transport\cite{nazarov2009quantum}. Improving the quality of electrostatic gating is an ongoing effort in the pursuit of ever-cleaner quantum transport.
In the study of two-dimensional (2D) van der Waals (vdW) materials, this demand for disorder reduction has led to the development of all-vdW gates\cite{rhodes_disorder_2019}, which employ hexagonal boron nitride (hBN) as a dielectric layer and few-layer graphite (FLG) as a gate electrode.
Thanks to the low charge trap density and absence of dangling bonds of hBN\cite{dean_boron_2010,meric_graphene_2010,mayorov_micrometer-scale_2011}, the enhanced disorder screening of FLG\cite{ponomarenko_tunable_2011}, and the overall atomic flatness of an all-vdW heterostructure, this approach has enabled substantial improvements in the resulting device quality across a range of vdW materials \cite{hunt_massive_2013,li_quantum_2016,zibrov2017tunable,yankowitz_tuning_2019,choi_electronic_2019}.
Some effort has been made to expand this technique to the gating of non-vdW materials, with improved transport quality reported in InSb\cite{kammhuber_conductance_2016,jekat_exfoliated_2020} and InAs\cite{elalaily_gate-controlled_2021} nanowires when hBN is used with non-vdW gate electrodes, and when FLG is used along with conventional thermal oxide dielectrics\cite{shani2024diffusive}.
The potential of all-vdW gating of non-vdW materials, however, remains under-explored \cite{zhang_fabrication_2022,Liang_2024}, and the expansion of this technique to the few-modes, quantum transport regime in quasi-one-dimensional (1D) systems has to date remained an open challenge.

Here we address this challenge by developing an all-vdW gated semiconductor device platform and employing it to study transport in a novel non-vdW material: an InSb nanoribbon (NR).
This NR belongs to the broader class of low-dimensional semiconducting materials with large spin-orbit interaction (SOI)~\cite{van_weperen_spin-orbit_2015}, large Land\'{e} $g$-factors \cite{nilsson_giant_2009,van_weperen_quantized_2013}, and relatively high mobility\cite{gul_towards_2015,badawy_high_2019,verma_high-mobility_2021}, such as InSb and InAs nanowires (NWs), which for more than a decade have underpinned research efforts ranging from topological superconductivity to spin-orbit qubits and spintronics.  \cite{nadj-perge_spinorbit_2010,nadj-perge_spectroscopy_2012,mourik_signatures_2012,pribiag2013electrical,van_den_berg_fast_2013,plissard_formation_2013,szombati2016josephson,gul_hard_2017,gul_ballistic_2018,yang_spin_2020,dvir2023realization}.
Compared to NWs, our NRs have a cross section that is much wider than it is tall, introducing anisotropy to the transverse quantum confinement.
Anisotropic confinement is expected to introduce a magnetic field angle dependence to both the $g$-factor and subband spacings in high-SOI materials \cite{pryor_lande_2006,schwan_anisotropy_2011,qu_quantized_2016,masuda_transport_2018,lei_gate-defined_2021,mu2020measurements}.
Such angle dependence, if coexistent with 1D ballistic transport in a NR, could allow for the decoupling with field angle of the semiconductor and superconductor Zeeman energies in hybrid devices for topological superconductivity, or enhance the tunability of spintronics devices.
However, the quantum transport properties of high-SOI NRs have not been studied, and neither ballistic, few-modes transport nor anisotropic field response have been realized.
Additionally, high-SOI NR (and NW) devices represent an ideal candidate for exploring the potential of all-vdW gating. This is because they are in most respects an excellent platform for realizing highly-sought physics, such as Majorana zero modes (MZMs)\cite{kitaev_unpaired_2001,lutchyn_majorana_2010,oreg_helical_2010}, yet studies in this field to date have been limited by disorder \cite{pan_quantized_2021,ahn_estimating_2021,frolov2023smoking,chen2019ubiquitous,das_sarma_disorder-induced_2021}, with some of this disorder being potentially introduced by the conventional metal-dielectric gate structures that have been typically used.

Using our all-vdW gated InSb NRs, we report in this work both the first successful realization of ballistic quantum transport in a non vdW-material with an all-vdW gate, and the first study of the transport properties of high-SOI NRs, measured across channel lengths of 1.5 $\mu$m, 400 nm, and 200 nm.
In all channels, we observe very low to near-zero gate hysteresis, consistent with the reduced charge trap density of the all-vdW gate.
In the 200 and 400 nm channels we observe quantized conductance, indicating that the NRs act as ballistic 1D channels\cite{van_wees_quantized_1988,van_houten_quantum_1996}, and we demonstrate that this quantization survives even in magnetic fields of only a few hundred mT, indicative of low disorder throughout the fabricated NR devices.
Finally, we observe field-angle anisotropy of the subband splittings coexistent with 1D ballistic transport, establishing NRs as a promising platform offering enhanced flexibility in experiments which typically use NWs. 
In the main text of this Letter, we fully characterize a single NR, with our multi-segment device geometry establishing our three channels as independently-probable segments, testing the NR across a total length of more than 2.5 $\mu$m.
Full characterization data from an additional, lithographically identical, NR device on a separate all-vdW gate is presented in Supplementary Information Section S8, and largely reproduces the results in this Letter.

\begin{figure}
    \centering
    \includegraphics{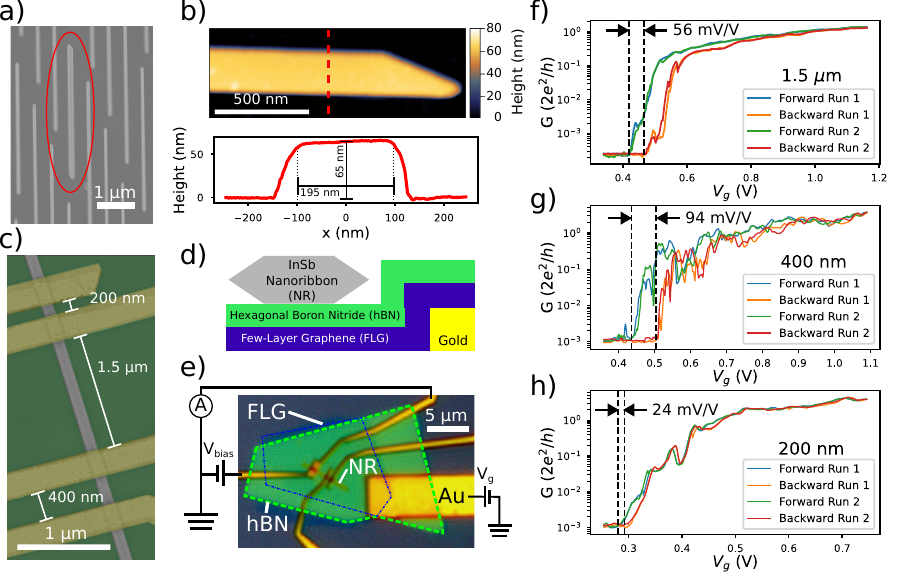}
    \caption{Summary and gate characterization of studied device. a) Scanning electron micrograph (SEM) of a typical nanowire growth field, with a nanoribbon (NR) circled in red. b) Atomic force microscope (AFM) scan of the NR used in our device, with a linecut taken at the dashed line. The NR's width is $\sim3$ times its height. c) False color SEM image of a lithographically identical device, showing the hBN dielectric (green), Ti/Au contacts (yellow), and NR (grey). Appropriate contact choice allows transport to be studied in uninterrupted channels of nominal length 200 nm, 400 nm, and 1.5 $\mu$m d) Schematic cross section of the studied device's all-van der Waals-material gate stack, contacted from below by a gold electrode. e) Optical image of the studied device, with a characteristic DC bias measurement schematically illustrated. f-h) Gate characterization in the 1.5 $\mu$m (f), 400 nm (g), and 200 nm (h) channels. Conductance, $\mathrm{G} = I_\mathrm{meas}/V_\mathrm{bias}$, is plotted in for two complete forward and backward loops of the gate voltage, $V_g$. The difference between forward and backwards threshold voltages is marked, expressed as a ratio with the total gate voltage swept ($\eta$ in the text). All channels show low hysteresis and high reproducibility, with the 200 nm channel showing near-zero hysteresis for most values of $V_g$. $V_\mathrm{bias}$ is 5 mV in (f), and 1 mV in (g-h).}
    \label{device_figure}
\end{figure}

The NRs used in our devices are grown by metal-organic vapor phase epitaxy alongside symmetric, ultra-thin NWs\cite{badawy_high_2019,shani2024diffusive}, as shown in Figure \ref{device_figure}a.
The symmetry-breaking planar expansion of the NRs during growth is likely a consequence of slow, layer-by-layer lateral accumulation occurring alongside the conventional vertical, gold-catalyzed VLS growth, as discussed in Ref.~\citenum{gazibegovic_bottom-up_2019}.
For the NR studied in this Letter, AFM (Figure \ref{device_figure}b) and SEM scans reveal it to have a width of $\sim$ 195 nm, and a height of $\sim$ 65 nm, giving a cross sectional aspect ratio of 3:1. 
A similar cross-sectional aspect ratio is measured for the additional device in the Supplemental Information.
Thus, our NRs can be understood as an intermediate between a sheet-like 2D nanoflake, which occurs when the lateral growth is comparable in rate to the VLS growth,\cite{zhang_fabrication_2022,gazibegovic_bottom-up_2019,verma_high-mobility_2021,rossi2023merging}, and the typical quasi-1D, hexagonally symmetric, NW, which occurs when the lateral growth rate is zero.
Importantly, since the NR growth is still VLS-dominated, the NR long axis, and thus transport, lies in the [-111] crystal axis, so that the NR has vanishing Dresselhaus\cite{dresselhaus1955} and large Rashba\cite{rashba1959} SOC\cite{van_weperen_spin-orbit_2015}. 

Our all-vdW gates, schematically illustrated in Figure \ref{device_figure}d, were assembled from commercially sourced (hqGraphene) bulk hBN and natural graphite which has been mechanically exfoliated\cite{huang_reliable_2015} and confirmed by AFM to be clean, atomically flat, and suitably thick (16.7 nm hBN, 0.95 nm FLG for the device in this letter). 
The stack was assembled via a conventional dry transfer technique\cite{zomer_fast_2014,purdie2018cleaning} and deposited onto a substrate prepatterned with Ti/Au (10 nm/90 nm) alignment markers and backgate contact electrodes, such that the electrode touched a corner of the FLG from underneath.
The sample was cleaned in hot chloroform to remove polymer residue and vacuum annealed at 400\degree C for several hours to further clean the stack surface and remove interfacial bubbles.
The NRs were transferred to this gate using a micromanipulator\cite{flohr_manipulating_2011}, ensuring the NRs sit entirely on the atomically flat portion of the gate, away from the Au electrode.
Finally, contacts, patterned by electron beam lithography, were deposited (10 nm/120 nm Ti/Au) after a short \textit{in-situ} Ar ion mill to remove the native oxide at the NR surface.
Full fabrication details can be found in Supplementary Information Section S1.
An optical micrograph of the device studied in this Letter, along with a characteristic measurement schematic, is presented in Figure \ref{device_figure}e.
A scanning electron micrograph of the second, lithographically identical device presented in the Supplemental Information is shown in Figure \ref{device_figure}c, showing our contact spacing scheme.
Measurements were performed in a dilution refrigerator with a base temperature of $\sim$ 8 mK using either DC or conventional low-frequency lock-in techniques, as indicated.

To understand the impact of our all-vdW scheme, we first characterize hysteresis and reproducibility in the NR gate response across all channel lengths.
For each channel length, the corresponding contacts are DC-biased with $V_{\textrm{bias}}$ (5 mV for the 1.5 $\mu$m channel, 1 mV for the 200 and 400 nm channels), and current, $I_\mathrm{meas}$, is measured while the gate voltage, $V_g$, is twice swept forwards and backwards, without resetting to zero between direction changes.
The resulting conductance, $\mathrm{G}= I_{\mathrm{meas}}/V_\mathrm{bias}$, is plotted for each channel in Figure \ref{device_figure}f-h.
To enable comparison across channels, we quantify hysteresis as the ratio of the difference between forward and backward threshold voltages (marked with dashed lines in Figure \ref{device_figure}f-h) to the total range of gate voltage swept through.
We find near-pinchoff hysteresis values of $56 ~\mathrm{mV/V},~94 ~\mathrm{mV/V}$ and $24~\mathrm{mV/V}$ for the 1.5 $\mu$m, 400 nm, and 200 nm channels, respectively, with the 200 nm hysteresis approaching zero for $V_g \gtrsim 0.35$ V.
These values match or outperform hysteresis values in optimized InSb devices on traditional oxide dielectrics\cite{gul_towards_2015}, and are comparable to values reported in previous InSb devices with hBN dielectrics\cite{jekat_exfoliated_2020,zhang_fabrication_2022}.
This hysteresis behavior was also found to be largely independent of the gate sweep rate, as discussed in Supplementary Information Section S2.
Additionally, in all channels, we see that the conductance traces, including fluctuations, are reproducible in both sweep directions, indicating that the induced electrostatic environment of our vdW gate and our disorder configuration are stable in time, with minimal temporally-fluctuating charge traps.
This combination of low hysteresis and good reproducibility across multiple channel lengths is consistent with the improved electrostatic gating expected from all our all-vdW approach.

\begin{figure}
    \centering
    \includegraphics{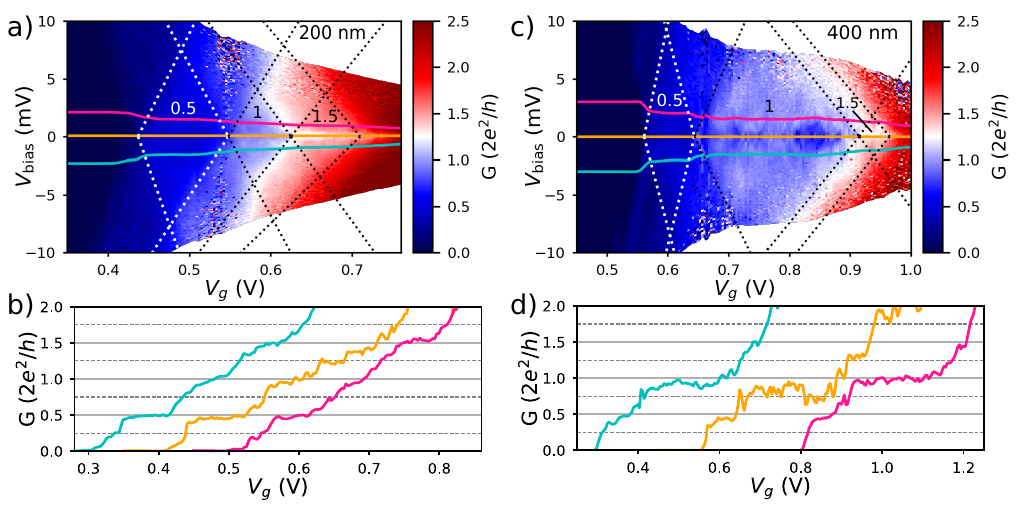}
    \caption{Quantized conductance in a 2.8 T out of plane magnetic field. a) Differential conductance, $G=\mathrm{d}I_\mathrm{meas}/\mathrm{d}V_\mathrm{bias}$ colormap as a function of $V_\mathrm{bias}$ and $V_g$ in the 200 nm channel. Diamond-shaped plateaus, marked with dashed lines to guide the eye, are visible at values of 0.5, 1.0 and 1.5 $G_0$, corresponding to the quantized, ballistic, quasi-1D transport through the first three Zeeman-split subbands in the NR. The dashed lines have been aligned with the plateau edges by following lines of high transconductance, $\partial G/\partial V_g$. b) Linecuts from a) at zero (orange) and finite (fuchsia and cyan) biases. For clarity, the positive and negative bias linecuts are offset in $V_g$ by +100 mV and -100 mV respectively. c) Differential conductance, $\mathrm{G}$, as a function of $V_\mathrm{bias}$ and $V_g$ in the 400 nm channel. The first three Zeeman-split plateaus are again visible, indicating quantized conductance. d) Zero and finite bias linecuts from c), with the positive and negative bias linecuts offset in $V_g$ by +120 mV and -120 mV, respectively.}
    \label{quant_figure}
\end{figure}

To further probe the quality of our short (200 nm and 400 nm) channels, we apply an out-of plane magnetic field, $B_z=2.8$ T, and search for quantized conductance. Because quantized resistance is easily spoiled by even small numbers of scattering events in the confined region, it constitutes a good benchmark of the electronic quality of the completed device and of the potential for using this nanofabrication approach more broadly in experiments requiring few-modes ballistic transport over the inter-contact distance.
Using standard lock-in techniques, with an excitation voltage of 50 $\mu$V at 19 Hz, we measure differential conductance $G = \mathrm{d}I_\mathrm{meas}/\mathrm{d}V_\mathrm{bias}$ as function of $V_\mathrm{bias}$ and $V_g$.
The resulting conductance colormaps are shown in Figure \ref{quant_figure}a for the 200 nm channel, and Figure \ref{quant_figure}c for the 400 nm channel, with linecuts in the $V_g$ axis at zero and finite bias shown in Figures \ref{quant_figure}b and d.
In both channels, clearly defined plateaus, visible as diamonds of uniform conductance in the colormaps (Figures \ref{quant_figure}a,c) and steps in the linecuts (Figures \ref{quant_figure}b,d) emerge.
These plateaus align with half integer values of the conductance quantum, $G_0=2e^2/h$, consistent with Zeeman splitting of the transverse quantum modes.
Additionally, in both channels, we observe intermediate-valued plateaus ($G \sim$ 0.25, 0.75 $G_0$), at finite $V_\mathrm{bias}$, as expected in a clean ballistic 1D system when the source and drain electrode chemical potentials lie in different subbands\cite{kouwenhoven1989_nonlinear}.
Together, these quantized and intermediate, finite-bias plateaus convincingly demonstrate that, at this field, our NR hosts ballistic, quasi-1D transport through individual Zeeman-split, spin-resolved subbands\cite{van_wees_quantized_1988,van_houten_quantum_1996} at both the 200 and 400 nm length scales.

We note that a finite series resistance, $R_S$, of 12.3 k$\Omega$ (13.8 k$\Omega$) was subtracted from the measured conductance in the 200 (400) nm channel, which includes the contact resistance of the device and the resistance of the filters and current amplifier (8.7 k$\Omega$ total) in our measurement setup.
upon subtraction of the known measurement resistance, these $R_S$ values correspond to contact interface resistances of 3.6 and 5.1 k$\Omega$, respectively. 
These $R_S$ values were determined by aligning the $G=1.0 G_0$ plateau in the zero bias linecuts of Figure \ref{quant_figure}b,d with the quantized value, and are comparable to $R_S$ extracted by fitting the full pinch-off of the 1.5 $\mu$m channel (see Supplementary Information Section S3 for further discussion).

The quantized plateaus can be used to determine the subband spacings, and thus the $g$-factor, in both ballistic NR channels.
The diamond plateaus in Figure \ref{quant_figure} close for values of $V_\mathrm{bias}$ which place the source and drain electrochemical potentials at the bottom of successive subbands, providing a direct measure of the subband splitting in the NR\cite{van_weperen_quantized_2013}.
At $B_z=2.8$ T, the lowest two subbands in each channel are expected to be the Zeeman-split transverse orbital ground state, so the height of the 0.5 $G_0$ plateau, $V_{0.5}$, is set by the Zeeman energy, $E_z = |E_{1\downarrow}-E_{1\uparrow}| = |g|\mu_B|\mathbf{B}| = eV_{0.5}$.
For the 200 nm channel, we can read off a height of $V_{0.5}\sim 8.2$ mV, corresponding to $|g| \sim 51$.
For the 400 nm channel, we can read off $V_{0.5} \sim 9.2$ mV, giving $|g|\sim 57$. 
Both of these values are reasonably consistent with the $g$-factor of conventional InSb NW \cite{van_weperen_quantized_2013,nilsson_giant_2009,badawy_high_2019,kammhuber_conductance_2016,kammhuber_conductance_2017,estrada_saldana_split-channel_2018}, suggesting that InSb NRs are relevant for MZM and spintronic applications where large $g$-factors are required.

Quantization in quasi-1D InSb NWs with comparable channel lengths to our 400 nm channel is not commonly reported in the literature, with quantization reports typically occurring in channels $\sim$200 nm in length\cite{van_weperen_quantized_2013,kammhuber_conductance_2016, badawy_high_2019}.
Our ballistic transport in the 400 nm channel is therefore indicative of a mean free path at least comparable to the best devices previously reported.
Additionally, the 2.8 T external magnetic field we have applied is comparable to or smaller than those typically used to observe quantization in long-channel devices\cite{kammhuber_conductance_2017}.
This magnetic field suppresses backscattering events which destroy quantization in geometrically defined 1D channels such as NWs and NRs\cite{van_houten_four-terminal_1988,buttiker_absence_1988,van_weperen_quantized_2013,kammhuber_conductance_2016}, so the need for a smaller field indicates relatively low disorder in our channel, which is consistent with high quality ballistic transport performance in our all-vdW gated NR device.
Further, while quantized conductance in InSb NW devices with channel lengths at or below 200 nm and conventional gating is more widely reported\cite{kammhuber_conductance_2016,van_weperen_quantized_2013,badawy_high_2019,estrada_saldana_split-channel_2018,verma_high-mobility_2021}, the data from the 200 nm channel of our NR device is exceptionally clean, consistent with the low degree of disorder suggested by the near-zero hysteresis in Figure \ref{device_figure}h.

\begin{figure}
    \centering
    \includegraphics{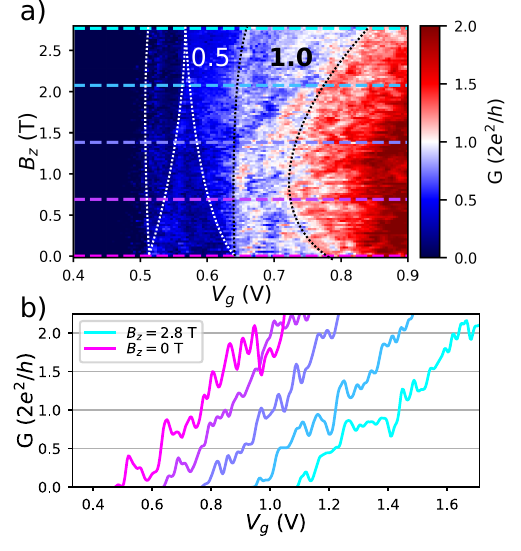}
    \caption{Out-of-plane field sweep, 400 nm channel. a) Differential conductance at $V_\mathrm{bias}=0$ as a function of $V_g$ and out-of-plane magnetic field, $B_z$. Both the 0.5 and 1.0 $G_0$ plateaus persist and continuously evolve through the full range of field values. A spin-split resonance is present before the onset of the 0.5 plateau, and is likely a signature of an incidental quantum dot in the wire. Dashed lines mark feature edges as a guide to the eye, and are placed by following lines of high  transconductance, $\partial\mathrm{G}/\partial V_g$. b) Linecuts from a), taken with a uniform spacing of 700 mT, offset in $V_g$ by 150 mV, starting from $B_z=0$.}
    \label{robust_figure}
\end{figure}

To further test the performance of our device, we study the quantization in our 400 nm channel at zero bias as the external out-of-plane magnetic field, $B_z$, is reduced. 
The resulting data  is plotted in Figure \ref{robust_figure}a, with dashed lines added to the plateau edges to guide the eye.
Linecuts at evenly spaced fields are shown in Figure \ref{robust_figure}c.
Data from the 200 nm channel can be found in the Supplementary Information Section S5.
In the 400 nm channel, the 0.5 and 1.0 plateau appear to persist and continuously evolve, both in the colormap in Figure \ref{robust_figure}a and the linecuts in Figure \ref{robust_figure}c, through the full range of the field sweep. 
The width of the 0.5 plateau in $V_g$ monotonically decreases as $B_z$, and thus $E_z$, decreases, but only disappears for $B_z = 0$, when the spin degeneracy is restored.
A similarly Zeeman-split feature appears before the onset of the 0.5 plateau, which we attribute to a spinful resonance of an accidental quantum dot in the channel\cite{wang2019crossover,gupta2024evidence}.
The 1.0 $G_0$ plateau also varies in $V_g$ width, though non-monotonically.
As the field is reduced from $B_z=2.8$ T, it first narrows until $B_z \sim 1$ T, then widens, ultimately persisting at $B_z=0$. 
As the 400 nm channel represents the edge of typical mean free path estimates in InSb NW, its quantization serves as a qualitative probe for disorder; the continuous evolution and persistence of plateaus as the backscattering-suppressing field goes to zero thus indicates low disorder in our NR device, potentially a consequence of the vdW gate.

Ultimately, we refrain from claiming zero-field quantization, as for low fields, fluctuations in $G$ become more prominent, preserving the plateaus but rendering the quantization increasingly imprecise.
This could be due to defects at the NR contact interface introduced by the ion milling of the native oxide prior to contact deposition\cite{gul_hard_2017}, which act as disorder and cause the contact to act non-Ohmically at low bias for certain values of $V_g$.
The resulting aperiodic suppression of $\mathrm{G}$ can be seen prominently in Figure \ref{quant_figure}c, where several regions of $G <1.0\ G_0$ are visible at low bias throughout the 1.0 plateau, and in Figure \ref{quant_figure}d, where we observe fluctuations in $G$ at $V_\mathrm{bias}=0$ which are greatly reduced at finite bias.
The $V_g$ values where these suppressions occur vary with field, creating ``scars'' in the 1.0 plateau in the field scan, as are visible in Figure \ref{robust_figure}a, which resemble previously reported disorder-induced features\cite{estrada_saldana_split-channel_2018} that can mimic re-entrant signatures of the helical liquid phase in high-SOI materials\cite{streda_antisymmetric_2003,pershin_effect_2004,kammhuber_conductance_2017}.
Further discussion of possible contact damage and resistance can be found in Supplementary Information Section S4.

\begin{figure}
    \centering
    \includegraphics{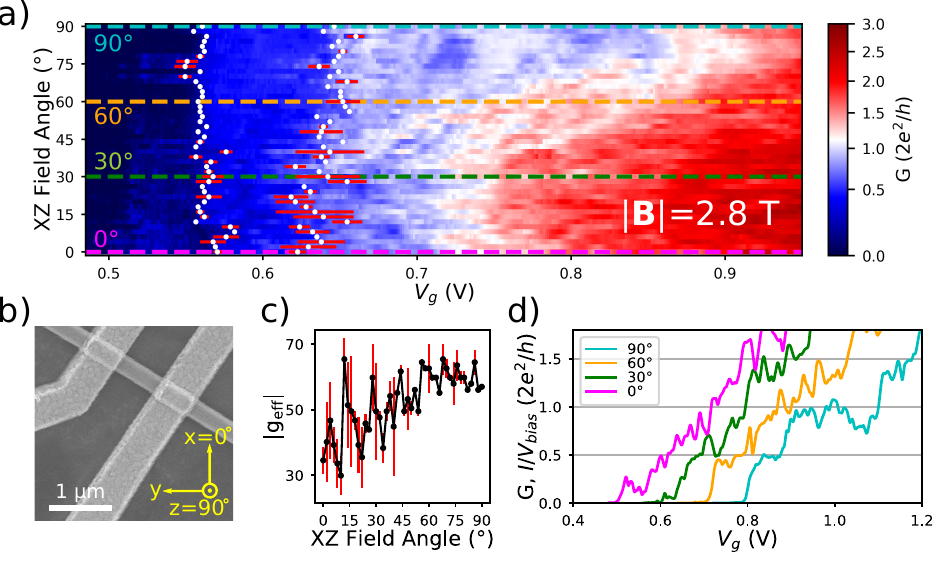}
    \caption{Field rotation from out-of-plane ($z$-axis, $\theta=90\degree$) to in-plane ($x$-axis, $\theta=0\degree$), at $|\mathbf{B}|=2.8$ T, in the 400 nm channel. a) Linear conductance, $\mathrm{G} = I_\mathrm{meas}/V_\mathrm{bias}$, at fixed $V_\mathrm{bias}=1$ mV as a function of field angle and $V_g$. Both the 0.5 and 1.0 plateaus have their width in $V_g$ modulated by the field angle, a consequence of anistropic confinement from the NR cross-section partially suppressing orbital contributions to both the $g$-factor and subband spacing as the field angle is reduced from $90\degree$. The edges of the 0.5 plateau are marked with white dots, corresponding to the location of the maximum value of transconductance near the visually identified plateau edge. Error bars capture approximate uncertainty in this edge identification.  b) SEM image of a lithographically identical device showing the coordinate system used for defining field angles. The studied NR lies in the $xy$-plane and makes an angle of 59$\degree$ with the $x$-axis. c) Effective $g$-factor magnitude, $|\mathrm{g_{eff}}|$, as a function of field angle, calculated from the $V_g$-width of the 0.5 plateau. $|\mathrm{g_{eff}}|$ is reduced by $\sim25\%$ as the field angle is brought from $90\degree$ to $0\degree$. Error bars are propagated from those of a). d) Linecuts from b) at equally spaced field angles, offset in $V_g$ by 80 mV, starting from $0\degree$.}
    \label{rotation_figure}
\end{figure}
                            
To probe for anisotropic magnetic field response in our NR, we study the dependence of the quantized conductance on the angle of the applied field.
In order to do so, we return to $|\textbf{B}| = 2.8$ T, and rotate the magnetic field from fully out of the NR plane (i.e. from the $z$-axis, $\theta=90\degree$) to fully in-plane ($x$-axis, $\theta=0\degree$) while applying a fixed $V_\mathrm{bias}=1$ mV across the 400 nm channel.
The linear conductance, calculated as $G=I_\mathrm{meas}/V_\mathrm{bias}$, is plotted in Figure \ref{rotation_figure}b, with linecuts in Figure \ref{rotation_figure}d.
Our plateaus persist through all angles of this field sweep, but we observe more pronounced fluctuations in the conductance for angles approaching zero, as seen in the linecuts.
This is likely because the NR forms an angle of $\sim 59\degree$ with the $x$-axis, resulting in an increasing portion of the field lying along the NR axis as the field angle is reduced, decreasing the efficacy of back-scattering suppression\cite{buttiker_absence_1988}. 
Additionally, the suppression of the conductance at low bias becomes more pronounced at low field angles, such that the contiguous 1.0 plateau region has $G <1.0\ G_0$ for $\theta \lesssim 30\degree$.
Nevertheless, we can still identify anisotropy in the field response.
The 0.5 plateau is widest in $V_g$ for $\theta \sim 90\degree$, and subtly narrows with reducing field angle.
To quantify this variation, we use the maximum of the transconductance, $\partial \mathrm{G}/\partial V_g$, to identify the edges of this plateau, indicated by the white dots superimposed on Figure \ref{rotation_figure}a.
Assuming the ratio of the width of the 0.5 plateau in $V_g$ to its height in $V_\mathrm{bias}$ to be angle-independent (see Supplementary Information Section S6 for further discussion), we use the previously performed bias spectroscopy at $\theta = 90\degree$ (Figure \ref{quant_figure}) to convert this measured plateau width into an effective $g$-factor, $g_\mathrm{eff}$, as a function of field angle, plotted in Figure \ref{rotation_figure}c.
The resulting $g_\mathrm{eff}$ is largest for field angles near $90\degree$, and is reduced by $\sim 25\%$ as the field angle approaches zero.
Anisotropic variation in the $V_g$-width of the 1.0 plateau is much more pronounced, with an approximately threefold reduction in width as the field angle varies from $90\degree$ to $0\degree$.
Error bars in Figure \ref{rotation_figure}a and c capture the uncertainty in identification of the plateau edges, and thus width, due to the previously mentioned mesoscopic conductance fluctuations.
A more detailed discussion of the determination of this uncertainty can be found in Supplementary Information Section S7.

To understand the observed anisotropy in both the 0.5 and 1.0 plateaus, we return to the intrinsic anisotropy of the NR.
When the field is perpendicular to the NR, the confinement length relevant for field-induced orbital motion - the $\sim195$ nm NR width - is much larger the electron gyroradius at $2.8 T$ ($\sim 20$ nm).
This enables both the enhancement of the $g$-factor via SOI, as in bulk InSb, and also the increase of the subband spacing $E_2-E_1$ by orbital effects\cite{van_wees_quantized_1988,berggren1988depopulation,qu_quantized_2016}.
As the field is made increasingly parallel to the NR, the relevant confinement length decreases, approaching the NR height, $\sim 65$ nm, as $\theta$ approaches $0$.
This partially suppresses orbital motion in the NR, reducing both the $g$-factor and enhancement of the subband spacing.
Such anisotropic field response is common in both quantum dots in gate-defined quantum dots based on InAs\cite{pryor_lande_2006} and InSb\cite{mu2020measurements} NWs, and in ballistic quantum point contacts (QPCs) on InSb two-dimensional electron gasses (2DEGs)\cite{qu_quantized_2016,masuda_transport_2018,lei_gate-defined_2021}. In contrast, here the anisotropic response arises naturally from the as-grown quasi-1D confinement, and is not due to gate-induced confinement.

In conclusion, we have demonstrated a technique for gating non-vdW materials using an all-vdW gate, and have used this gating scheme to perform the first transport characterization of VLS-grown InSb nanoribbons, a novel high-SOI, quantum-confined material.
With this platform, we have demonstrated remarkably clean transport in the NRs, including the first realization of few-modes, ballistic quantum transport in a non-vdW material with an all-vdW gate.
In particular, we have shown low to near-zero gate hysteresis across length scales of 200 nm, 400 nm, and 1.5 $\mu$m, and quantized conductance plateaus in the 200 and 400 nm channels, which qualitatively persists with decreasing external magnetic field to fields as low as a few hundred mT.
This behavior is largely reproduced in a lithographically identical second device, detailed in Supplementary Information Section S8.
These results, and their reproducibility, suggest that the all-vdW gate is effective at reducing disorder when compared to conventional oxide-based gates, and definitively establish NRs as quasi-1D objects, with a mean free path comparable to the best reported InSb NWs.
Additionally, we have shown that the Zeeman and subband splittings in the NR studied in this Letter are anisotropic with the angle of the applied field, owing to the anisotropic transverse confinement and strong SOI in the NR. 
To our knowledge, such anisotropy is absent from ballistic III-V nanowires.

Our results have several implications. On a broad level, the all-vdW material gate scheme described in this letter is a flexible platform, and can be immediately applied beyond nanoribbons to almost any non-vdW material where extremely high quality, low-hysteresis gating is desired.
The specific case of InSb nanoribbons gated with an all-vdW stack, as we have demonstrated, holds promise as a platform for realizing sought-after phenomena that rely on the ability to fabricate low-disorder gate-tunable ballistic quantum devices in strong-SOI 1D systems.
For example, the low-field, long-channel quantized conductance which we have reported is highly conducive to proposed helical liquid experiments\cite{streda_antisymmetric_2003,pershin_effect_2004,rainis_conductance_2014}, and with further refinement, the InSb NR/vdW gate platform may aid in definitive observation of the helical liquid state, which would be a major milestone to the conclusive realization of Majorana zero modes.
Additionally, the $g$-factor anisotropy in a NR may enable further flexibility in Majorana device design, allowing the decoupling of Zeeman energy from the superconducting gap without introducing the large Dresselahus contributions\cite{qu_quantized_2016} and possible gate instabilities\cite{lei_gate-defined_2021} endemic to existing anisotropic InSb 2DEG QPCs.
These benefits are especially relevant to current efforts in scalable selective-area-growth methods\cite{lee2019_inas_sag,aseev_ballistic_2019,jung2022pbte_sag}, where the full control of aspect ratio allows for the creation of arbitrary NR networks with variable cross-sections, compatible with MZM braiding proposals\cite{alicea_non-abelian_2011}.

\section*{Acknowledgements}
All aspects of the work at UMN were supported by the Department of Energy under Award No. DE-SC0019274. Portions of this work were conducted in the Minnesota Nano Center, which is supported by the National Science Foundation through the National Nano Coordinated Infrastructure Network (NNCI) under Award Number ECCS-1542202. Eindhoven University of Technology acknowledges the research program “Materials for the Quantum Age” (QuMat) for financial support. This program (registration number 024.005.006) is part of the Gravitation program financed by the Dutch Ministry of Education, Culture and Science (OCW). Eindhoven University of Technology acknowledges European Research Council (ERC TOCINA 834290).
We would also like to thank Zach Anderson, Chiou Yang Tan, and Martin Greven at UMN for allowing us to use their vacuum furnace, and providing assistance with the annealing step of device fabrication.

\bibliography{references}
\section*{Author Contributions}
C.J.R. and V.S.P. designed the experiments. C.J.R and T.L. fabricated the devices. C.J.R performed the measurements on all devices and analyzed the data. P.L., G.B., M.R., and E.P.A.M.B. grew the nanoribbons. C.J.R. prepared the manuscript, with input from all authors. V.S.P. and P.A.C. supervised the project.
\section*{Competing interests}
The authors declare no competing interests.
\end{document}

\maketitle

\section{Device Fabrication Details}
\subsection{Flake Exfoliation}
The 2D materials used in this letter were prepared using conventional mechanical exfoliation of bulk van der Waals (vdW) materials. Natural graphite purchased from hqGraphene was used to prepare the few-layer graphite (FLG) used for our gate electrode, and commerically available hexagonal boron nitride (hBN), also purchased from hqGraphene, was used to prepare our gate dielectric. In order to minimize adhesive residue of the top surface of our hBN flakes, blue surface protection tape (Nitto SPV-224PR-MJ) was used in place of the typical Scotch tape for the exfoliation of both materials.

For the deposition of exfoliated flakes, commercially available Si wafers with 285 nm of thermal oxide (NOVA Electronics HS39626-OX) were diced into 1 cm square chips, and prepared in accordance with the high-yield procedure described in Ref.~\citenum{huang_reliable_2015}. These chips were cleaned in N-Methyl-2-pyrrolidone for $\sim$12 hours, then sonicated in $\sim60\degree$ C acetone, then $\sim60\degree$ C isopropyl alcohol (IPA)for 5 minutes each, and dried with N2. The chip was then baked on a 180$\degree$ C hotplate for 5 minutes, and cleaned in oxygen plasma (PlasmaTherm Advanced Vacuum RIE, 100W, 99sccm O2, 100 mtorr) for 5 minutes. A second piece of blue tape was pressed into and removed from the previously prepared source tape to prepare a daughter tape of exfoliated flakes. This daughter tape was pressed onto this cleaned chip, and the chip and tape are baked on a 100$\degree$ C hotplate for two minutes before gentle pressure on the tape was applied with a cotton swab. The chip and tape were then allowed to cool to room temperature, and the tape was slowly peeled off the chip. This process was repeated to create both graphite and hBN source chips.

Candidate flakes were identified on each chip with optical microscopy, and were characterized by atomic force microscopy to ensure they were of suitable thickness, atomically flat, and free of contamination from tape residue.
\subsection{Backgate Chip Preparation}

An additional 1 cm square chip of the Si with 285 nm thermal oxide was prepared, and cleaned using the same NMP, acetone, and IPA sequence described above. A trilayer resist stack was spun onto the chip, consisting of two layers of PMMA 495K A4, and a top layer of PMMA 950k A2, all spun at 4000 rpm for 60 s and baked for 5 min at 180$\degree$ C between layers. Local alignment markers, bonding pads, and the gold electrode for making electrical contact to the FLG were patterned with electron beam lithography (EBL), using a Vistec 5000+ HR electron beam pattern generator operating at 100 kV. The patterned sample was developed for 80 s in a 3:1 mix of IPA and methyl isobutyl ketone (MIBK), rinsed with IPA, and descummed for 15 s in a PlasmaTherm Advanced Vacuum reactive ion etcher, using oxygen plasma at 30 W, 40 sccm, and 100 mTorr vacuum. 10/90 nm of Ti/Au was then evaporated onto the chip using a CHA e-beam evaporator (1 \AA/s for Ti, 3\AA/s for Au), and liftoff was performed in hot acetone, aided by sonication.
\subsection{Stack Preparation}
After suitable candidate flakes were identified, they were assembled into an hBN/FLG heterostructure using a modified version of the polycarbonate (PC) film pickup technique described in Ref. \citenum{zomer_fast_2014}. A glass slide with PC film (6$\%$ solid PC dissolved in chloroform, spread into a think layer on separate slide and allowed to dry) stretched over a small piece of PDMS was placed into the micromanipulator of a commercially available hqGraphene manual transfer station. An optically clean portion of the PC film was identified under the transfer station's microscope, and used to pick up first the hBN, and then the FLG flakes. For both flakes, the PDMS/PC film stamp was aligned with the flake, and lowered into contact with the vdW source chip until just before the PC film had contacted the target flake. The transfer station stage was then slowly heated to 100$\degree$ C, such that thermal expansion brought the PC film fully into contact with the flake, and drove out particularly mobile interfacial bubbles\cite{purdie2018cleaning}. The stage was then cooled and the slide lifted from the source chip, with the target flake now adhered to either the PC film (hBN) or the PC film and hBN picked up in the first step (FLG). After the heterostructure was assembled in this way, the stack was deposited onto the prepatterned backgate chip by lowering the slide until the entirety of the PDMS stamp was in contact with the chip. The stack was aligned such that the gold backgate electrode touched a portion of the FLG, but left a considerable flat region of undisturbed hBN/FLG on which to fabricate our device. The transfer station stage was then heated to $\sim180\degree$ C as rapidly as possible, melting the PC film off the slide and allowing for the slide and PDMS stamp to be removed. Finally, the stack was cleaned of residual PC film by soaking in a bath of hot chloroform for $\sim2$ hours. After removal from the chloroform, the sample was rinsed in IPA, and blown dry with N2.
\subsection{Device Fabrication}
Following stack preparation, the sample was annealed in a quartz tube furnace at 400$\degree$ C for $\sim$4 hours in order to remove interfacial bubbles and further burn off organic contaminants. After annealing, the sample was allowed to cool under vacuum for an additional 90 minutes before being removed. A nanoribbon (or nanowire) was then deposited on each stack using a micromanipulator setup installed on an optical microscope, and optical micrographs of the stack were taken for accurate contact design and alignment. Contacts were then written with EBL, using the same A4/A4/A2 resist stack as for the backgate deposition. The patterned sample was descummed in oxygen plasma for 15 s, using the same parameters as before, and loaded into the CHA e-ebeam evaporation system. After vacuum was established, but prior to deposition, the sample was milled for 3 minutes using a KRI Ar ion mill (10.7 sccm, 200V discharge, 3.0 A discharge, 3.0 A emission), and vacuum was re-established. Ti/Au (10/120 nm) was then deposited at a rate of 1 \AA/s, and liftoff was performed in hot acetone, followed by an IPA rinse. 

\section{Impact of Sweep Rate on Gate Hysteresis}
\begin{figure}
    \centering
    \includegraphics{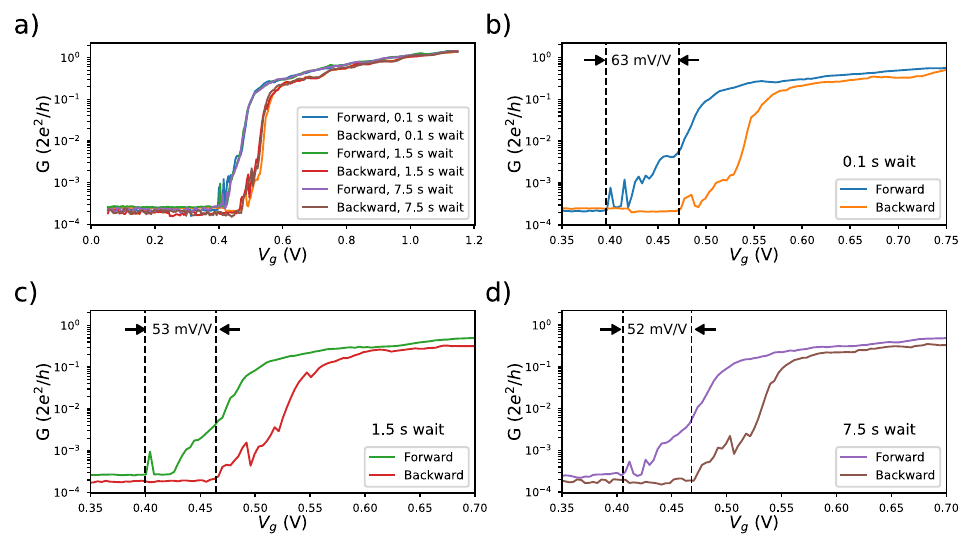}
    \caption{Hysteresis in the 1.5 $\mu$m channel as a function of wait time between the setting of gate voltage, $V_g$, and the measurement of current through the device. a) Linear conductance, $G=I_\mathrm{meas}/V_\mathrm{bias}$, for forward and backwards sweeps of $V_g$ with wait times of 0.1, 1.5, and 7.5 s. The impact of wait time on the observed hysteresis is negligible, and the traces lie almost entirely on top of each other. b-d) Details of the hysteresis loops in the neighborhood of the threshold voltage for each wait time. Threshold voltages for each sweep are indicated with horizontal dashed lines, and the corresponding normalized near-pinchoff hysteresis value is given, using the same definition as in the main text.}
    \label{sleep_figure}
\end{figure}
In general, transient charging effects mean that the rate at which a gate is swept in a semiconducting device may impact the hysteresis observed between forward and backwards sweeps.
To gauge the presence of this rate-dependent, we repeated the threshold hysteresis analysis performed in the main text for a variety of effective sweep rates of the gate voltage, $V_g$, focusing on the 1.5 $\mu$m channel. 
As our gate voltage source (QuTech IVVI Rack S1h HV source) does not have true slew rate control, we approximated changes to the sweep rate by altering the time between the $V_g$ being set and the current though the device being measured. 
Because the steps between each $V_g$ setting are very small (4 mV) relative to the voltage range swept through (1.2 V), we expect this approximation to be sufficient for qualitative exploration of sweep rate effects.

Figure \ref{sleep_figure} shows the results of this study, with forward and backward runs plotted for wait times of 0.1 s, 1.5 s, and 7.5 s.
While some reduction in hysteresis is observed as the wait time is increased from 0.1 s to 1.5 s, the magnitude of the reduction is rather small, amounting to a change of only about 16\%.
Additionally, no significant change in hysteresis is observed as the wait time is increased from 1.5 s to 7.5 s.
Therefore, we conclude that transient, sweep-rate related effects are not a meaningful contributor to the performance of our device, and that the hysteresis data reported in the main text is representative of the intrinsic performance of the device.

\section{Mobility and Contact Resistance in 1.5 $\mu$m channel}
\label{mobility_section}

\begin{figure}
    \centering
    \includegraphics{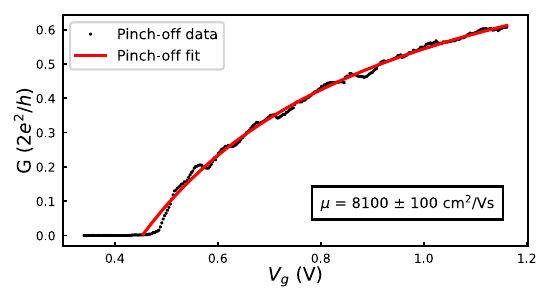}
    \caption{Forward pinch-off and fitting for mobility extraction in the 1.5 $\mu$m channel. A fixed bias of 5 mV is used, and conductance is calculated as G $=I_\mathrm{meas}/V_\mathrm{bias} $. No series resistance is subtracted, as it is a variable to be fitted.}
    \label{mobility_figure}
\end{figure}

In the 1.5 $\mu$m channel, we expect transport to be diffusive, dominated by scattering events within the NR.
We can thus model the NR as a channel of length $L$ with fixed mobility, $\mu$, and carrier density controlled by capacitive coupling, $C$, to the gate electrode, with voltage $V_g$.
Combined with the fixed series resistance in our measurement, $R_S$, which captures the lump sum of the contact resistance at the NR/contact interface, the resistance of our filtered lines, and the resistance of the transimpedance amplifier used to measure current, we can quantitatively describe the $V_g$-dependent conductivity of the NR as 
\begin{equation}
    G(V_g)=\bigg(R_S +\frac{L^2}{\mu C(V_g-V_{th})} \bigg)^{-1},
    \label{pinchoff_eqn}
\end{equation}
where $V_{th}$ is the threshold voltage of the NR.

To quantitatively apply this model, we first model the NR (treated as a metallic object) and gate stack cross section in the COMSOL MultiPhysics software, and use a simple 2D Laplace solver to determine a capacitance of 292 aF/$\mu$m.
We can then fit the forward $V_g$ sweep in the 1.5 $\mu$m channel to Equation \ref{pinchoff_eqn} to determine both the mobility and series resistance of the NR. 
This fit, plotted in Figure \ref{mobility_figure}, gives us a mobility of $\mu = 8100\pm100$ cm$^2$/Vs, and a series resistance $R_S = 12.2$  k$\Omega$. 
The total resistance of our filters and measurement is known to be 8.7 k$\Omega$, allowing us to identify a contact resistance of 3.5 k$\Omega$.

This mobility is comparable to, but somewhat lower than, some previously reported values\cite{gul_towards_2015,badawy_high_2019}.
However, we argue that the exact value of mobility is an imperfect measure of the quality of transport in our device.
After all, fitting to Equation \ref{pinchoff_eqn} depends heavily on a characterization of the capacitance, $C$. 
While more elaborate methods to numerically simulate devices and derive their capacitance exist, such as Schr\:{o}dinger-Poisson techniques, even these fail to capture the actual intricacies of a real, nanofabricated device.
Additionally, the fitting procedure itself is fraught with potential errors.
For example, overly strict adherence to fitting certain parts of the pinch-off trace, such as the steepest parts of the beginning of the curve, can lead to overestimation of $\mu$\cite{mcculloch_avoid_2016,choi_critical_2018}.
Finally, mobility measurements characterize only the diffusive, long-channel performance of a device, while the spintronic and topological applications which make InSb NRs appealing require operation in the ballistic, few-modes regime.
Therefore, because our mobility is of the same order as in other reports, we believe the high quality of quantization, over longer-than-typical length scales, which we report in the main text is the best available metric to establish that our device is operating in a regime of relatively clean quantum transport

\section{Contact Damage and Low-Bias Device Performance}
\label{contact_damage_section}

In order to make electrical contact to an InSb NR (or NW), the few nanometers of native oxide present at the semiconductor surface must first be removed.
For all devices reported in this letter, we have accomplished this task using a brief \textit{in-situ} Ar ion mill of the NR in our e-beam evaporator prior to contact deposition.
While this successfully removes the oxide, it also has the potential to damage the InSb surface at the contact area, leaving a rough surface behind\cite{gul_hard_2017} and creating significant disorder near the metal/semiconductor interface.
Further, neither the degree to which this damage occurs, nor the impact of the particular disorder realization on transport, is guaranteed to be the same at each contact.
As a result, the transport through each channel studied in our devices is heavily impacted by the particular contacts used to establish said channel.

\begin{figure}
    \centering
    \includegraphics{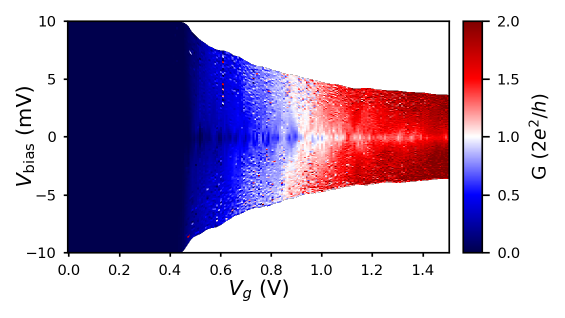}
    \caption{Differential conductance, $G=\mathrm{d}I_\mathrm{meas}/\mathrm{d}V_\mathrm{bias}$, colormap as a function of $V_\mathrm{bias}$ and $V_g$ in the 1.5 $\mu$m channel, with zero applied magnetic field. No evidence of quantized conductance is present, as this channel is in the diffusive transport regime, but the effects of a somewhat non-Ohmic contact are visible at low biases throughout the whole $V_g$ range. Here a series resistance of 12.2 k$\Omega$ has been subtracted, as determined in the mobility fitting.}
    \label{contact_damage_figure}
\end{figure}

The variation in contact resistances and transport performance found for each device can be understood as a consequence of variable contact quality. 
After subtracting measurement resistances, the resistances attributable to the contact interface for the 200 nm, 400 nm, and 1.5 $\mu$m channels are 3.6, 5.1, and 3.5 k$\Omega$, respectively.
Already, this suggests that at least one of the contacts to the 400 nm channel is more problematic than average.
This damage is further evidenced by the increased oscillation in, and general suppression of, the low-bias conductance in the 400 nm channel, as reported in the main text.
We see similar low-bias behavior in a $V_\mathrm{bias}$ vs. $V_g$ conductance scan of the 1.5 $\mu$m channel, shown in Figure \ref{contact_damage_figure}; the effect is particularly pronounced for $0.6$ V $\lesssim V_g \lesssim$ 1.0 V, as in the 400 nm channel.
As this behavior is not prominently observed in the 200 nm channel, we can conclude that, at minimum, the contact shared by the 400 nm and 1.5 $\mu$m channel was especially damaged during ion milling, resulting in non-Ohmic, disordered behavior at low bias.
Thus, the shared 400 nm/1.5 $\mu$m contact acts as a sort of weak link for determining device performance in both channels, and illustrates the importance of high-yield techniques for achieving high-quality contacts in future experiments.
This may also explain why both the highest quality quantization and lowest hysteresis were observed in the 200 nm channel, as its contacts appear to simply be less disordered.

Literature reports suggest that oxide removal and surface passivation of the InSb prior to contact formation with an ammonium polysulfide solution can improve device performance over ion milling\cite{kammhuber_conductance_2016,gul_hard_2017}, but we ultimately found this process difficult to implement in our fabrication process.
However, the idea is well-supported, and we are optimistic that future explorations of all-vdW gates on InSb NWs or NRs may show even further enhancements to transport quality with successful implementation of the sulfur passivation approach.

\section{200 nm Channel Magnetic Field Sweep Data}
\label{main_text_200nm_sweep}

\begin{figure}[H]
    \centering
    \includegraphics{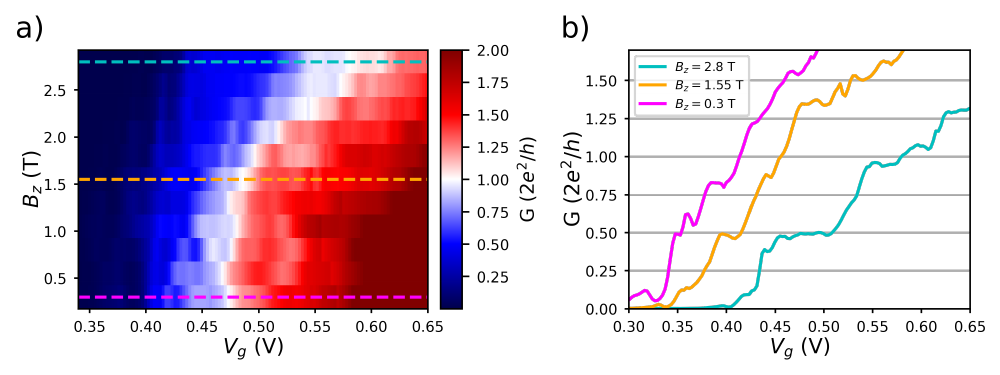}
    \caption{Out-of-plane field sweep, 200 nm channel. a) Linear conductance at $V_\mathrm{bias}=1$ mV as a function of $V_g$ and out-of-plane magnetic field, $B_z$. Both the 0.5 and 1.0 $G_0$ plateaus appear to persist and continuously evolve through the full range of field values. b) Linecuts from a) at various field values, offset in $V_g$ by -60 mV (1.55 T) and -120 mV (0.3 T) for clarity. In all traces, plateau-like features with intermediate, non-half-integer values of $G$ are observed in addition to the quantized plateaus. The exact nature of these features is unclear, and may vary with field strength.}
    \label{200nm_sweep_figure}
\end{figure}

As with the 400 nm channel in the main text, we can view the evolution of the quantization in the 200 nm channel as the applied out-of-plane magnetic field, $B_z$, is reduced as a test of device performance.
The resulting data are plotted in Figure \ref{200nm_sweep_figure}.
Like in the 400 nm channel, we observe a smooth, monotonic narrowing of the 0.5 $G_0$ plateau as the field is decreased, corresponding to a reduction in the Zeeman splitting of the first subband.
We also observe a more or less continuous evolution of the 1.0 $G_0$ plateau down to our lowest studied field, $B_z = 300$ mT.
The continuity of these two features is, much like in the 400 nm channel, suggestive of high quality, few mode quantum transport which can survive in the absence of large backscattering-suppressing magnetic fields, and is consistent with the interpretation of low overall disorder in our device.
However, we again cannot make a claim of zero-field quantization, as we unfortunately did not collect data for $B_z < 300$ mT.

Interestingly, unlike in the 400 nm channel, we do not see a substantial expansion of the width of the 1.0 $G_0$ plateau at large ($\gtrsim 1.5$ T) values $B_z$.
In the 400 nm channel, we believe this expansion is a consequence of the width of the NR allowing orbital effects at large out-of-plane fields to increase the splitting between the first and second transverse modes.
Therefore, we can interpret the lack of this expansion in the 200 nm channel as an indication that the short channel length serves to suppress these orbital effects, even if the NR is wide.
This suppression may also be a consequence of screening or interfacial effects from the metallic contacts.

Additionally, we note the presence of plateau-type features at intermediate quantized values (e.g G $\sim 0.75 \ G_0$), especially in the low-field region connected to the high-field 1.0 $G_0$ plateau. 
For the high-field intermediate plateaus, such as the 1.25 $G_0$ feature observed in the $B_z = 2.8$ T linecut, this is likely an artifact of the finite bias (1 mV) used for this sweep, which will give rise to such intermediate plateaus if the electrochemical potentials of each contact lie in different subbands.
At lower fields, however, it is unclear whether these features are a finite bias effect, or a consequence of some disorder and increased scattering rendering the quantization somewhat imprecise, as observed in the 400 nm channel in the main text.

\section{Lever Arm in the 400 nm Channel at Additional Magnetic Field Angles}

\begin{figure}[H]
    \centering
    \includegraphics{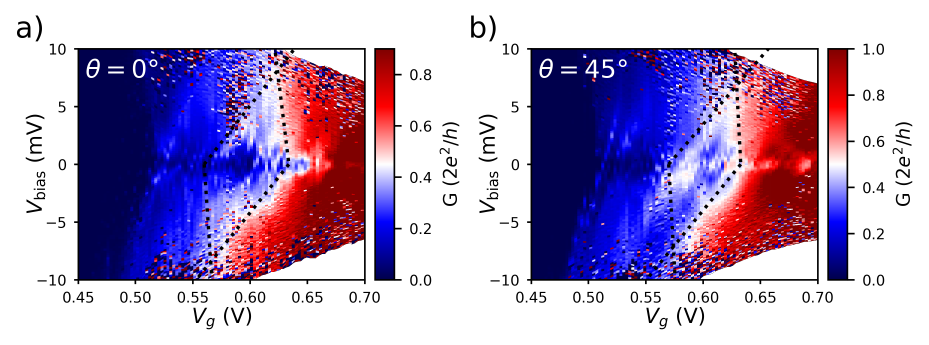}
    \caption{Differential conductance, $G=\mathrm{d}I_\mathrm{meas}/\mathrm{d}V_\mathrm{bias}$, colormap as a function of $V_\mathrm{bias}$ and $V_g$ in the 400 nm channel for different angles of the external 2.8 T magnetic field. a) Field angle of $\theta=0$, fully in-plane with the NR. b) Field angle of $\theta=45\degree$, halfway between in-plane and out-of-plane. In both panels, the edge of the 0.5 $G_0$ plateau has been identified by hand and indicated with black dashed lines. Both field angles give a lever arm of $\sim 110$ meV/V, the same as was found for the $\theta=90\degree$ (fully out-of plane) case, justifying the use of plateau width as a proxy for $g$-factor in the main text.}
    \label{lever_arm_figure}
\end{figure}

In order to test our assumption in the main text that the lever arm, the ratio of a plateau's width in $V_g$ to its height in $V_\mathrm{bias}$, is independent of the angle of the applied magnetic field, we performed bias spectroscopy of the 400 nm channel with a 2.8 T external magnetic field applied fully in-plane and at $45\degree$ between in- and out- of plane ($0\degree$ and $45\degree$ in the convention used in the main text, respectively).
The resulting data are plotted in Figure \ref{lever_arm_figure}, and include the manually identified and  edges of the 0.5 $G_0$ plateau at each angle, overlayed as black dotted lines.

Using these edges, we can determine that, for both field angles, the lever arm is $\sim 110$ meV/V, as in the 90$\degree$ case reported in the main text.
Compared to the $\theta=90\degree$ case, the quality of quantization for both angles is notably worse, with the $\theta=45\degree$ data being the cleaner of the two.
This is consistent with the observed increase in conductance fluctuations and overall suppression observed as the field is rotated from out-of-plane (90$\degree$ to in-plane (0$\degree$), and is likely a consequence of the magnetic field being less effective at suppressing backscattering events, as may be caused by the previously discussed contact interface damage, the more in-plane it becomes.
This degradation of quality makes the exact identification of the plateau edges difficult, and the large fluctuations in conductance necessitate that the edges be identified visually, as opposed to using the more typical lines of high transconductance ($\partial G/\partial V_g$)\cite{van_weperen_quantized_2013}.
However, we note that for both angles these fluctuations largely share the same slope as other $V_g$-dependent features, and so while the overall height and width of the plateaus may be somewhat uncertain, their ratio, and thus the lever arm, should be still well-established.
We therefore conclude that our use in the main text of the plateau  width to determining the anisotropy of the $g$-factor is reasonable and justified.
\section{Uncertainty in Anisotropic $g$-Factor}

\begin{figure}[H]
    \centering
    \includegraphics{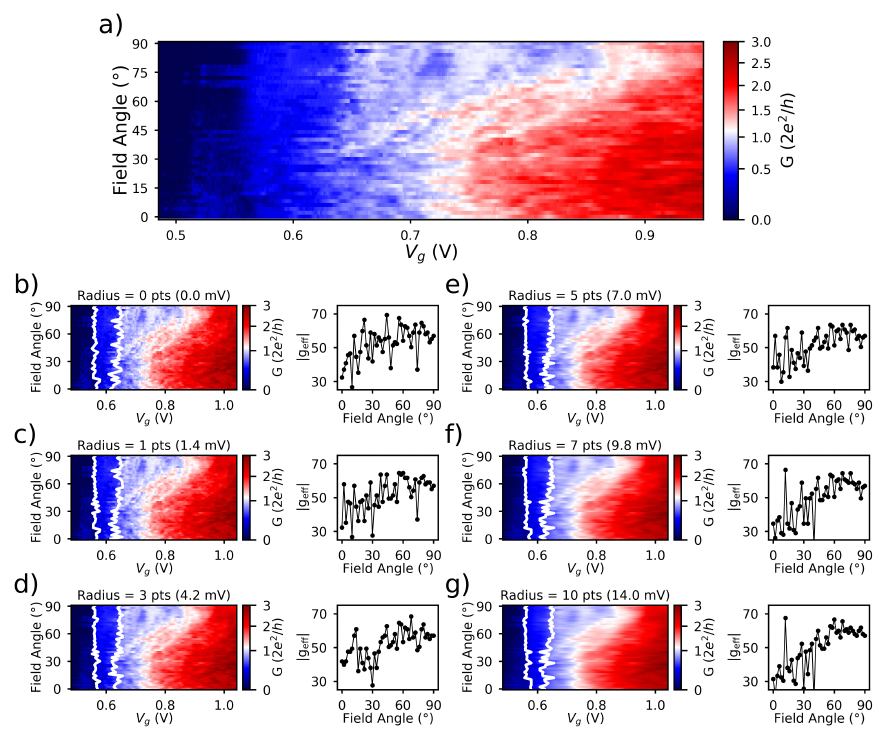}
    \caption{Determination of uncertainty for plateau edge identification in magnetic field rotation data. a) Unannotated and unfiltered data, as presented in Figure 4a of the main text. b-g) Plateau edges and $g$-factor anisotropies for varying degrees of filtering, as used to identify the uncertainty in the $g$-factor anisotropy in Figures 4a and 4c of the main text. For each panel, the linear conductance data shown in the colormap has been filtered along the $V_g$-axis with a Gaussian filter of the indicated radius (with the corresponding value in mV of $V_g$ noted). The identified plateau edges are plotted as white lines on the colormap, and the corresponding angle-dependent $g$-factor, identified from the plateau width as described in the main text, is plotted.}
    \label{uncertainty_figure}
\end{figure}

Identification of the edges of the 0.5 $G_0$ and 1.0 $G_0$ plateaus in the field angle sweep of Figure 4a in the main text is made difficult by the oscillations in conductance caused by mesoscopic scattering events.
As we have discussed, this scattering, whether it originates from contact interface damage or otherwise, renders quantization imperfect, and oscillations in $G$ can obscure the true edge of a plateau.
This difficulty in edge identification means it is difficult, if not impossible, to exactly determine the width of the 0.5 $G_0$ plateau in $V_g$, as we must do in order to extract the corresponding $g$-factor for main text Figure 4c.
Furthermore, as these oscillations which obscure our plateau edges are a consequence of reproducible scattering events, we cannot simply repeatedly measure the plateau edges to get a statistical average and uncertainty of the plateau width.

One solution to this difficulty is to filter the data slightly, running the linear conductance data of Figure 4a through a Gaussian filter along the $V_g$ axis only.
In principle, this smooths out small oscillations around the plateau value, leaving the edges of the plateau unchanged (though softer).
One can then use the maximum of the transconductance ($\partial G/\partial V_g$) in the optically identified $V_g$ region of the transition between plateaus as a consistent proxy for the plateau edge.
Such an optical identification of the search region is necessary to avoid false edge identifications due to any remaining oscillations of $G$ within the plateau itself, which may also have a large transconductance despite not being an actual edge.
Repeating this process for both the start and end of the 0.5 $G_0$ plateau at each angle gives the width of the plateau, which we can then convert into an angle-dependent $g$-factor using the previously performed bias spectroscopy for a fully out-of-plane 2.8 T magnetic field (main text Figure 2c), as described in the main text.

In practice, the location of the edges, and thus the plateau width and $g$-factor, found using this method varies with the radius of the filter used.
After all, wider filters give less noise overall, but allow a point to be impacted by ever further-away oscillations in $G$, shifting the exact location of the transconductance maxima.
Narrower filters, on the other hand, leave more noise in the data, allowing the transconductance maxima to be shifted from the true edge by relatively smaller oscillations in the transition region.
This effect can be seen in Figure \ref{uncertainty_figure}, where we have plotted the filtered data, plateau edges, and resulting $g$-factor vs angle traces at a variety of filter radii.
For all filter radii, the trend of the $g$-factor decreasing as the field angle is swept from 90$\degree$ to 0$\degree$ is preserved, but the exact value of the $g$-factor at a particular angle may vary as the filter radius is changed.
The degree of this variance is what we have used to determine the uncertainty of our edge identification, and thus our $g$-factor anisotropy.
Specifically, for each field angle, we have found the plateau edge corresponding to each integer filter radius (measured in data points, each of which is 1.4 mV of $V_g$) in the set [0,10], inclusive.
We have used the average of these edges as the overall plateau edge, and the standard deviation of the edge values as the error.
These are the values indicated in main text Figure 4a,c.

In this way, the error corresponds to a sort of ``uncertainty in analysis''.
If the plateau edge is largely unchanged as the filter radius is varied, then that edge can be considered to robustly identified, with minimal uncertainty.
On the other hand, if the plateau edge location varies significantly with the filter radius, then the actual location of the edge is much less certain, heavily obscured by nearby oscillations which give comparable transconductance values.
While this is not a statistical uncertainty, it does effectively capture the difficulty in edge identification.
Further, our results in using this method indicate that the reported $g$-factor anisotropy is genuine, though the exact shape of the curve is not known with a high degree of certainty. 
We note that a true statistical analysis would require many thermal cyclings of our dilution refrigerator in order to sample different possible disorder configurations of the device, and is beyond the scope of this Letter.

\section{Additional Devices}

Two additional devices were fabricated on the same chip as the device reported in the main text, and cooled down in the same cycle for measurement.
Each device was lithographically identical to the one in the main text, with the same configuration of a 1.5 $\mu$m channel for diffusive transport, and 200 nm and 400 nm channels for ballistic transport.

As in the main text, both of these devices were fabricated upon NRs, with similar flattened cross-sections.
This can be seen in the AFM scan and linecut of SD1's 1.5 $\mu$m channel in Figure \ref{dev1_1.5um_figure}a.
Again, the NR is short, only $\sim$52 nm tall, but rather wide.
Unlike the one in the main text, this NR is not symmetric, and instead has a substantial slope on one side. 
This slope obfuscates a clear identification of the NR width; taken at the top facet, this NR is $\sim$ 135 nm wide, while taken halfway down the slope, it is $\sim$ 175 nm wide.
Further, this slope suggests the growth of these NRs, while likely still a consequence of the simultaneous lateral and vertical growth mechanisms discussed in the main text, may not be entirely uniform from ribbon to ribbon (or that both NR have slopes, and one is just upside-down with respect to the other).
Notably, both this and the main text NR were selected from the edges of a growth field intended to produce conventional, $\sim 60$ nm diameter NWs, suggesting that the different growth dynamics at the edge of a large growth patch may be an important factor in determining NR vs NW growth outcomes.

One of these devices was partially destroyed by electrostatic discharge during either wirebonding or loading, and as a result did not provide useful data.
The other device, Supplemental Device 1 (SD1), survived the loading process, and is characterized in the following section.
For expediency, all measurements on SD1 were performed using conventional DC techniques, i.e. without the use of a lock-in amplifier.

\subsection{1.5 $\mu$m Channel}

\begin{figure}[H]
    \centering
    \includegraphics{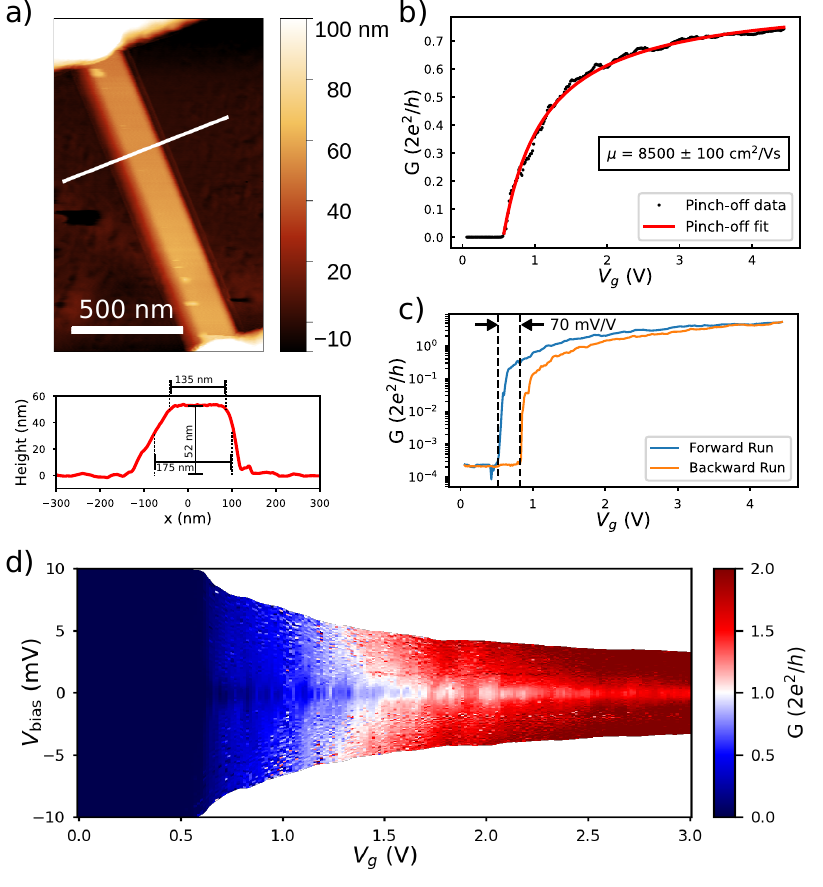}
    \caption{Characterization and 1.5 $\mu$m channel transport of Supplemental Device 1 (SD1). a) Atomic force microscope (AFM) image of the 1.5 $\mu$m channel, with the plotted linecut taken at the position indicated by the white line. As in the main text, this device is fabricated on a nanoribbon, much wider (135+ nm) than it is tall (52 nm). We note that this ribbon has a strong slant on its left side, making exact width identification difficult. b) Linear conductance vs $V_g$ of the 1.5 $\mu$m channel at a fixed 5mV bias, as well as a fit to Equation \ref{pinchoff_eqn}, with $C$ = 202 aF/$\mu$m. The extracted mobility is comparable to the main text device, but the series resistance of 15 k$\Omega$, corresponding to a contact resistance of 6.3 k$\Omega$, is much higher than in any channel of the main text device. c) Hysteresis in the linear conductance of the 1.5 $\mu$m channel, at a fixed 5 mV bias. d) Differential conductance as a function of $V_\mathrm{bias}$ and $V_g$ in the 1.5$\mu$m channel at zero external field. Conductance suppression and oscillations are notable at low bias throughout the $V_g$ range, consistent with the damaged contact interfaces suggested by the high contact resistance found in fitting. For c) and d), the fitted series resistance of 15 k$\Omega$ has been subtracted.}
    \label{dev1_1.5um_figure}
\end{figure}

As with the device in the main text, we can use the diffusive transport in the 1.5 $\mu$m channel of SD1 as a benchmark for gate performance, contact damage, and transport quality.

For example, we can again perform a pinchoff measurement, sweeping $V_g$ while applying a fixed 5 mV bias, and fit the data to Equation \ref{pinchoff_eqn} to extract mobility and contact resistance.
Here we have again used COMSOL to determine our device capacitance, with the measured NR cross section and hBN thickness (25.7 nm) yielding a capacitance of 202 aF/$\mu$m.
The pinchoff data, and its corresponding fit, are shown in Figure \ref{dev1_1.5um_figure}b, and give a mobility of $\mu \sim 8500\pm 100$ cm$^2$/Vs, a value comparable to, if slightly higher than was found for the main text device in Section \ref{mobility_section}.
However, the fitted series resistance, $R_S$ = 15 k$\Omega$, is much higher that for the main text device.
In particular, once known measurement resistances ($8.7$ k$\Omega$) are removed, this corresponds to an actual contact interface resistance of 6.3 k$\Omega$, much higher than any of the contact resistances in the device in the main text.
This suggests that more contact interface damage was likely induced in SD1, and further highlights the importance of optimizing this step in future experiments.
The impacts of this poor contact quality can be seen in a $V_\mathrm{bias}$ vs. $V_g$ conductance scan of the 1.5 $\mu$m channel, shown in Figure \ref{dev1_1.5um_figure}d. 
As in the main text device (see Section \ref{contact_damage_section}), heavy oscillation and suppression of conductance, especially at low $V_\mathrm{bias}$ is present, suggesting imperfectly-Ohmic contacts.

Despite this contact damage, the $V_g$-hysteresis of our device is still very low, as shown in Figure \ref{dev1_1.5um_figure}c. 
We measure a range-normalized threshold hysteresis of $\sim 70$ mV/V, comparable to the results in the main text. 
This may suggest that the low hysteresis is a consequence of the all-vdW gate, as it persists in both our studied devices even in channels with apparent contact damage.

\subsection{400 nm Channel}

We can next study the transport in the 400 nm channel, where, based on the results in the main text, we expect to be close to the limit of ballistic transport.

\label{SD1_400nm_section}
\begin{figure}[H]
    \centering
    \includegraphics{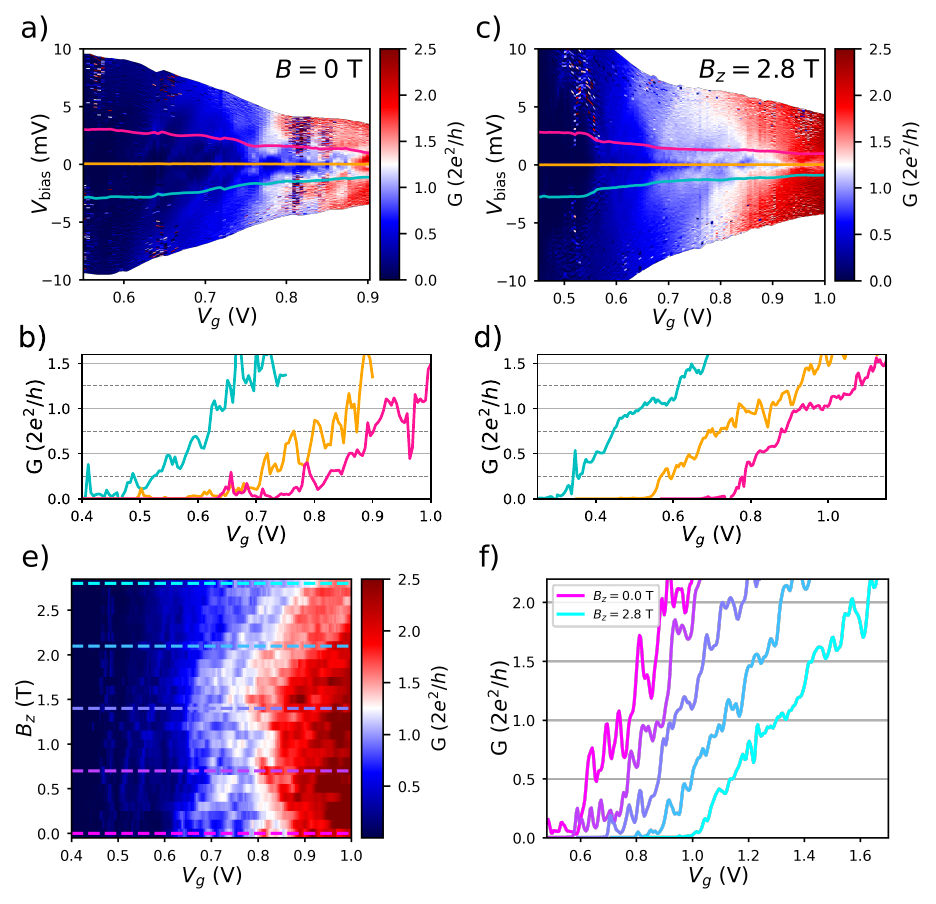}
    \caption{Transport in the 400 nm channel of SD1. a) Differential conductance, $G=\mathrm{d}I_\mathrm{meas}/\mathrm{d}V_\mathrm{bias}$, colormap as a function of $V_\mathrm{bias}$ and $V_g$ at zero external magnetic field. Hints of a quantized 1.0 $G_0$ plateau are present, but obscured by low bias suppression of the conductance. Additionally, Coulomb blockade resonances, indicative of an accidental quantum dot, are clearly observed for $V_g < \sim 0.75$ V. b) Linecuts from a) at zero (orange) and finite (fuchsia and cyan) biases. For clarity, the positive and negative bias linecuts are offset in $V_g$ by +150 mV and -150 mV respectively. c) Differential conductance as a function of $V_\mathrm{bias}$ and $V_g$ in an external, out-of-plane magnetic field, $B_z=2.8$ T. The first two Zeeman-split quantized plateaus are much more clearly identifiable, though some low-bias suppression remains. d) Zero and finite bias linecuts from c), with the positive and negative bias linecuts offset in $V_g$ by +220 mV and -220 mV, respectively. e) Linear conductance at $V_\mathrm{bias}=1$ mV as a function of $V_g$ and out-of-plane magnetic field, $B_z$. Both the 0.5 and 1.0 $G_0$ plateaus appear to persist and continuously evolve through the full range of field values, but the quality of quantization becomes worse as $B_z$ approaches zero. f) Linecuts from e), taken every 700 mT, each offset in $V_g$ by 120 mV, starting from $B_z = 0$. }
    \label{dev1_400nm_figure}
\end{figure} 

We begin with bias spectroscopy at zero external magnetic field, with differential conductance as a function of $V_\mathrm{bias}$ and $V_g$ plotted in Figure \ref{dev1_400nm_figure}a and linecuts in Figure \ref{dev1_400nm_figure}b.
As in the 1.5 $\mu$m channel, we observe heavy suppression of the conductance at low bias, as well as large oscillations, which especially dominate the zero-bias (orange) linecut in Figure \ref{dev1_400nm_figure}b.
Additionally, we see diamond-shaped resonances of higher-than-surrounding conductance for $V_g$ between 0.6 and 0.75 V.
These resonances are consistent with Coulomb blockade diamonds\cite{nazarov2009quantum}, and indicate an incidental quantum dot is present in our device\cite{gupta2024evidence}, most likely localized near the damaged region of the contact.
However, we do still see traces of few-mode quantum transport, even at zero field.
While heavily obscured by the contact damage scattering features, there does appear to be a diamond with $G\sim 1.0 \ G_0$ in Figure \ref{dev1_400nm_figure}a, and the finite-bias (cyan and fuchsia) linecuts which traverse this region do have hints of steps at 0.5 and 1.0 $G_0$, as would be expected for gate-tunable finite-bias transport through a single spin-degenerate subband.
However, as for the device in the main text, this evidence is not conclusive enough, especially in light of the heavy scattering present at low bias, to claim true zero-field quantization.

We do, however, observe quantized conductance when the external out-of-plane field, $B_z$, is increased to 2.8 T.
The resulting bias spectroscopy data, plotted in Figure \ref{dev1_400nm_figure}c-d, clearly shows the 0.5 and 1.0 $G_0$ Zeeman-split quantized plateaus.
We note that, while the quality of the data has been much improved, owing to the suppression of scattering by the external field, evidence of the damaged contacts still persists at low bias, and with Coulomb blockade resonances near $V_g \sim 0.5$ V.
Additionally, we can see that, like in the main text device, the 1.0 $G_0$ plateau in Figure \ref{dev1_400nm_figure}c is much wider in $V_g$ space than the 0.5 $G_0$ plateau, consistent with the enhanced splitting in energy of the first and second transverse quantized mode due to orbital effects enabled by the NR's width.

As in the main text, we can monitor the evolution of this quantization as $B_z$ is reduced in order to gauge its robustness.
The resulting data, plotted in Figure \ref{dev1_400nm_figure}e-f, show evolution which is qualitatively similar to that of Figure 3 in the main text.
It is important to note that this data, as in Section \ref{main_text_200nm_sweep}, is linear conductance data, taken at a fixed bias of 1 mV, meaning that intermediate quantized plateaus may be present 
As in previous field sweeps, we see here the narrowing of the Zeeman-induced 0.5 $G_0$ plateau with decreasing $B_z$. 
However, this feature does not vanish for $B_z=0$, originating at low fields from finite bias effects instead of Zeeman splitting.
We also observe a non-monotonic, but continuous, evolution of the 1.0 $G_0$ plateau, with a minimum width occurring for $B_z\sim 1.5$ T.
This non-monotonic evolution was also noted in the main text device, and we can attribute it to a competition in determining the spacing between the spin-up state of the ground state transverse mode and the spin-down state of the first excited transverse mode between Zeeman splitting (which shrinks the splitting with increasing $B_z$) and orbital effects (which increase the splitting). 
Like the modulation of the 1.0 $G_0$ plateau width with field angle in the main text, we take this as a consequence of the nanoribbon geometry enabling orbital effects, which the relatively long length of the 400 nm channel does not suppress. 
Throughout the field sweep, there are the large conductance oscillations expected for this device, which are most prominent for low fields, consistent with our zero-field bias spectroscopy.
We also see field-sensitive resonances for $V_g < 0.6$ V, again consistent with an incidental quantum dot.
Taken as a whole, this data suggests that signatures of few-mode quantum transport survives down to low or even zero field, but we again do not claim true quantization at zero field.

In all plots, a total series resistance of 14.5 k$\Omega$, corresponding to a contact interface resistance of 5.8 k$\Omega$, has been subtracted, determined by aligning the 1.0 $G_0$ features with their quantized values in Figure \ref{dev1_400nm_figure}d.

\subsection{200 nm Channel}
\begin{figure}[H]
    \centering
    \includegraphics{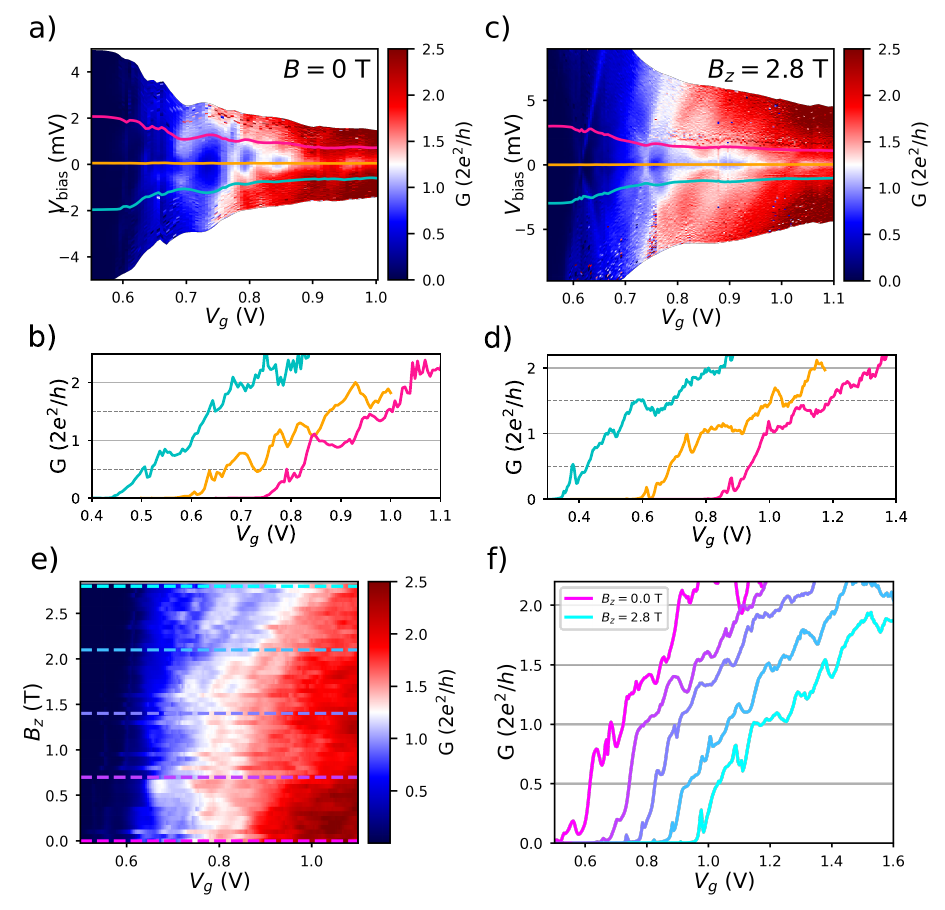}
    \caption{Transport in the 200 nm channel of SD1. a) Differential conductance, $G=\mathrm{d}I_\mathrm{meas}/\mathrm{d}V_\mathrm{bias}$, colormap as a function of $V_\mathrm{bias}$ and $V_g$ at zero external magnetic field. As in the 400 nm channel, hints of a quantized 1.0 $G_0$ plateau are present, but are even more heavily obscured by low bias suppression of the conductance. b) Linecuts from a) at zero (orange) and finite (fuchsia and cyan) biases. For clarity, the positive and negative bias linecuts are offset in $V_g$ by +150 mV and -150 mV respectively. In both finite bias linecuts, a 1.0 $G_0$ plateau is visible, as are the expected 0.5 and 1.5 $G_0$ finite bias intermediate plateaus. c) Differential conductance as a function of $V_\mathrm{bias}$ and $V_g$ in an external, out-of-plane magnetic field, $B_z=2.8$ T. The first two Zeeman-split quantized plateaus are somewhat more clearly identifiable, but a strong residual low-bias suppression and crisp Coulomb blockade resonances from another accidental quantum dot obscure them. d) Zero and finite bias linecuts from c), with the positive and negative bias linecuts offset in $V_g$ by +250 mV and -250 mV, respectively. e) Linear conductance at $V_\mathrm{bias}=1$ mV as a function of $V_g$ and out-of-plane magnetic field, $B_z$. Both the 0.5 and 1.0 $G_0$ plateaus appear to persist and continuously evolve through the full range of field values, but the quality of quantization becomes worse as $B_z$ approaches zero. f) Linecuts from e), taken every 700 mT, each offset in $V_g$ by 90 mV, starting from $B_z = 0$. }
    \label{dev1_200nm_figure}
\end{figure}

We can repeat the measurements and analysis of Section \ref{SD1_400nm_section} on the 200 nm channel of SD1, in order to evaluate transport when the channel length is solidly in the ballistic regime.
The resulting data are plotted in Figure \ref{dev1_200nm_figure}.

Given the quantization seen in the 400 nm channel, we might expect the data in our 200 nm channel to be even higher in quantization quality, as was the case in the device reported in the main text. 
However, that does not turn out to be the case.
Instead, the 200 nm channel appears to be even more impacted by contact interface damage than the 400 nm channel.
Looking first at the zero-field data in Figure \ref{dev1_200nm_figure}a-b, we again see strong hints of quantization at finite bias, both in the linecuts and the suggestion of a 1.0 $G_0$ diamond, but the severe suppression and oscillation of conductance at obscures this at low bias, and prevents us from claiming robust zero-field quantization.

Like in the 400 nm channel, at $B_z=2.8$ T, the data sharpens up, as can be seen in Figure \ref{dev1_200nm_figure}c-d.
However, the strong suppression at low bias persists, suggesting that the backscattering we attribute to contact damage is not as effectively suppressed at this field as in other channels.
Nevertheless, we can still identify the 0.5, 1.0 and even 1.5 $G_0$ plateaus as and steps in the linecuts, but clear identification of any diamonds beyond the 0.5 $G_0$ plateau is somewhat obscured by conductance suppression.
Strong Coulomb-blockade resonances of enhanced conductance also appear superimposed on the data, especially at $V_g \sim 0.6$ V and $\sim 0.75$ V, which are again indicative of the presence of an accidental quantum dot in our device.
We also note that the 1.0 $G_0$, so far as one can identify it, does not seems to persist for the same extent in $V_g$ as in the 400 nm channel.
This mirrors the behavior of the main text device, and is likely a consequence of the shorter channel, as discussed in Section \ref{main_text_200nm_sweep}.

Sweeping the magnitude of $B_z$, we see behavior consistent with the 200 nm channel in the main text device, as discussed in Section \ref{main_text_200nm_sweep}.
This data, which is again linear conductance taken at 1 mV bias, is presented in Figure \ref{dev1_200nm_figure}e-f.
In particular, the 0.5 and 1.0 $G_0$ plateaus evolve continuously throughout the full range of the field scan (with a 0.5 plateau appearing at $B_z=0$ T as a finite bias effect), but the 1.0 $G_0$ plateau is intersected and obscured by a field-sensitive ``scar'', as in the main text device's 400 nm channel.
This is consistent with the suspected contact damage.
Still, this data is strongly suggestive of few-mode quantum transport surviving to zero external field, but strong fluctuations in the conductance, especially at low bias, prevent us from claiming full zero-field quantization.

In all plots, a total series resistance of 12.9 k$\Omega$, corresponding to a contact interface resistance of 4.2 k$\Omega$, has been subtracted, determined by aligning the 1.0 $G_0$ features with their quantized values in Figure \ref{dev1_200nm_figure}d. 
This is lower than in either the 1.5 $\mu$m or 400 nm channels, hinting that the 200 nm contact shared with the 1.5 $\mu$m channel is responsible for most of the scattering issues, while the contact used only for the 200 nm channel is likely of reasonable quality, is not the dominant contributor to the contact resistance.
Nevertheless, this is still 600 $\Omega$ higher than the contact resistance of the main text device's 200 nm channel. 

\subsection{Conclusions From Supplemental Device 1}
On the whole, SD1 represents a second NR device which performs very similarly to the device reported in the main text, with the main difference being the final quality of the quantization involved.
Specifically, SD1 reproduces the low hysteresis in $V_g$ sweep, the quantization of conductance at field in both the 200 and 400 nm channels, persistence of the quantum few-mode effects to low magnetic fields, and suggestion of NR-specific orbital effects in the 400 nm channel of the main text device.
Indeed, the only phenomenon which we did not reproduce in SD1 was the field angle sweep, simply because such data was not collected during this chip's cooldown window.

The existence of this second, largely successful, device suggests several conclusions about our all-vdW gated NR platform.
Firstly, the presence of low gate hysteresis across both devices, despite their significant difference in contact quality, may indicate that this hysteresis is a consequence of the gating scheme, though this is still inconclusive without a control test of both NRs and NWs on other gating options, which is beyond the scope of this Letter.
Similarly, the fact that both NR devices showed evidence of orbital effects in their 400 nm channels suggests that the impact of the extra width of the NR cross section is meaningful when not suppressed by short (e.g. 200 nm) channel effects.
Finally, the degradation of quantization quality in SD1 compared to the main text device, coupled with the across the board higher contact resistances in SD1, reinforces the bottleneck nature of the contact interface in these devices for determining the ultimate quality of transport, singling it out as the next element to be improved in future implementations of our technique.

\bibliography{references}